\documentclass{aa}  
\usepackage{natbib}
\usepackage{graphicx}
\usepackage{txfonts}
\usepackage{multirow}
\usepackage[colorlinks=true,linkcolor=blue,citecolor=blue,urlcolor=blue]{hyperref}

\bibpunct{(}{)}{;}{a}{}{,} 

\begin{document} 

  \title{Importance of modelling the nebular continuum in galaxy spectra}
  
  \author{Henrique Miranda\inst{1,2}, Ciro Pappalardo\inst{1,2}, José Afonso\inst{1,2}, Polychronis Papaderos\inst{1,2,3}, Catarina Lobo\inst{3,4}, Ana Paulino-Afonso\inst{3}, Rodrigo Carvajal\inst{1,2}, Israel Matute\inst{1,2}, Patricio Lagos\inst{3}, and Davi Barbosa\inst{1,2}}
  \institute{Instituto de Astrof\'{i}sica e Ci\^{e}ncias do Espa\c{c}o, Universidade de Lisboa - OAL, Tapada da Ajuda, PT1349-018 Lisboa, Portugal
  \and
  Departamento de F\'{i}sica, Faculdade de Ci\^{e}ncias da Universidade de Lisboa, Edif\'{i}cio C8, Campo Grande, PT1749-016 Lisboa, Portugal
  \and
  Instituto de Astrof\'{i}sica e Ci\^{e}ncias do Espa\c{c}o, Universidade do Porto - CAUP, Rua das Estrelas, PT4150-762 Porto, Portugal
  \and
  Departamento de F\'{\i}sica e Astronomia, Faculdade de Ci\^encias, Universidade do Porto, Rua do Campo Alegre 687, PT4169-007 Porto, Portugal
  }


 
  \abstract
  {Neglecting to model stellar and nebular emission has been shown to have a significant impact on the derived physical properties of galaxies experiencing high levels of star formation. This impact has been seen at low redshifts for galaxies in a period of extremely significant star formation, the so-called extreme emission-line galaxies. It has also been suggested as a more general phenomenon among star-forming galaxies at high-redshifts. Even though various studies have approached the issue, a clear limit for the relevant effect of nebular contribution to the total optical emission has not been established.} 
  {We aim to correlate the nebular contribution in the optical regime with different tracers and to define a threshold, in terms of the nebular contribution, above which there is a significant impact on the estimation of physical properties of galaxies. Additionally, we want to investigate the implication of the results for high-redshift galaxies.}
  {We selected a sample of galaxies from SDSS-DR7 with a wide range of star-forming activity levels and analysed their spectra with two conceptually distinct spectral fitting tools: one that self-consistently models stellar and nebular emission and ensures that the best-fitting star formation and chemical enrichment history obtained reproduces the observed nebular characteristics of a galaxy (FADO), and another that lacks such a self-consistency concept (STARLIGHT) and was applied using a purely stellar base. We estimated the nebular contribution and correlate it with different tracers. Then, we compared the stellar properties estimated by the two spectral fitting tools for different degrees of optical nebular contribution. Additionally, we estimated the stellar properties using FADO in pure-stellar mode to further strengthen the robustness of our results.}
  {The rest-frame H$\alpha$ and H$\beta$ equivalent widths (EWs) show a strong linear correlation with the optical nebular contribution and are suitable tracers. We find that for an optical nebular contribution above 8\%, which corresponds to EW(H$\alpha$)$\simeq$500 \AA\, and EW(H$\beta$)$\simeq$110 \AA, there is a significant impact on the estimated physical properties and underlying stellar populations of a galaxy. Given the different definition of FADO for the continuum, this threshold actually corresponds to EW(H$\alpha$)$\simeq$375 \AA\, for works considering a pseudo-continuum, which is more commonly used in the literature. These findings were corroborated when considering the results from the application of FADO in pure-stellar mode. Considering the observed redshift evolution of EW(H$\alpha$), galaxies in the stellar mass range between M$_{*}$=10$^{7}$--10$^{11}$ M$_{\odot}$ will reach, on average, this threshold in the $z$$\sim$2--6 interval and the optical nebular contribution cannot be neglected.}
  {Our results highlight the importance of taking into account both stellar and nebular continuum when analysing the optical spectra of star-forming galaxies. In particular, this is a fundamental aspect for galaxies with a rest-frame EW(H$\alpha$)$\gtrsim$500 \AA\ (or the scaled value of 375\AA\ for pseudo-continuum measures). At low redshifts, this mostly impacts extreme emission line galaxies, while at higher redshifts it becomes a dominant aspect given the higher star-forming activity in the younger Universe. In light of current JWST observations and future instruments designed for high-redshift observations, such as MOONS, this reveals a critical issue that ought to be taken into consideration.}

  \titlerunning{To model or not to model: nebular continuum in galaxy spectra}
  \authorrunning{H. Miranda et al.}

  \maketitle
%

\section{Introduction}
\label{introduction}

Studying the importance of modelling the nebular emission from galaxies has been an active topic for a number of years \citep[e.g.][]{kru95,izo97,papa98,papa02,and03,gus07,scha09,rei10,atek11,papa12,card19,card22,guna20,ciro21,mir23}. In particular, its  aim has primarily been to analyse galaxies hosting very strong star-forming (SF) activity, the so-called extreme emission-line galaxies (EELGs). Among these systems, it has been shown that for blue compact dwarfs (BCDs) and HII galaxies, it is fundamental to model the stellar and nebular emission to retrieve accurate physical properties \citep{kru95,izo97,papa98,gus01,gus07}. In addition to BCDs, this analysis has also been applied to other SF luminous compact objects, such as green peas \citep{izo11}. On the modelling side, \cite{and03} updated previously existent evolutionary synthesis models \citep[from][]{schu02} by including both continuum and line emission from the nebular gas. They found that for the case of low metallicity and young ages, the nebular emission can contribute as much as 60\% to the observed broad-band fluxes.

The nebular contribution is highly sensitive to the age of the stellar population, showing a rapid decline as the population ages. For instance, several works have demonstrated that the equivalent width (EW) of H$\alpha$, a suggested tracer of the nebular contribution, decreases to relatively low levels around 10 Myr after a star formation burst \citep[e.g.][]{leit99,cid11,FADO,ciro21}. Consequently, stellar populations older than 10 Myr are expected to exhibit a relatively low nebular contribution. Beyond age, metallicity also plays a crucial role, being inversely related with the production of ionizing photons per unit mass \citep[e.g.][]{weil01,cid11}.

In recent decades, EELGs have been studied extensively from low-redshifts up to the epoch of reionisation \citep[e.g.][]{salz89,terl91,izo97,carda09,amo10,atek11,lag11,vdwel11,amo15,smit15,bru20,end21,breda22,boy22,simm23,ller24}. These systems are generally characterised by high specific star formation rates, low stellar masses, sub-solar metallicities, and small amounts of dust (noting that these properties are similar to those of local HII galaxies). One consequence of these properties are the high EWs of emission lines. In fact, H$\alpha$, H$\beta$ and [OIII]$\lambda$5007 EWs are commonly used to identify these systems and although there is no clear definition of an EW limit to identify EELGs, the considered limit varies in the range between hundreds and thousands of angstroms \citep[e.g.][]{carda09,vdwel11,amo15,end21,boy22,boy24}. It should be noted that although in most cases the ionising radiation leading to the high EWs of EELGs has its origin on star formation, it can also originate in active galactic nuclei \citep[AGNs, e.g.][]{carda09,dav23,ller24,boy24}. In the case of SF galaxies, the high EWs of EELGs point towards very prominent SF activity, which also implies a considerable nebular emission. In fact, for an EW(H$\alpha$)=1000 \AA, the nebular line emission leads to an enhancement of the $r$ band magnitude of $\sim$0.7 \citep{papa22}, thus highlighting the relevance of the nebular emission in the case of galaxies with such high EW values.

Galaxies hosting such a strong SF activity must have a significant nebular emission and carefully accounting for this emission arises as a fundamental aspect of their analysis. This was one of the motivations for the development of the spectral fitting code known as Fitting Analysis using Differential
evolution Optimization \citep[FADO;][]{FADO}. This is a population spectral synthesis code that self-consistently fits the optical spectrum of a galaxy, considering both the stellar and nebular continuum. This tool was used to investigate both stellar and nebular emission when deriving the physical properties of galaxies, using synthetic spectra \citep{card19,ciro21} and also real data \citep{breda22,card22,lag22,mir23,guo24}.

\cite{breda22} analysed a sample of $\sim$ 400 EELGs from SDSS-DR7 \citep{sdss,sdss-dr7} with the objective of studying their physical properties and the fundamental correlations between them. In addition to these two goals, this study also clearly highlighted the importance of incorporating nebular emission when modelling these highly SF objects.

\cite{mir23} examined how the additional consideration of the nebular emission affects the star-forming main sequence (SFMS). These authors used a sample of $\sim$180 000 SF galaxies from SDSS-DR7 with a mean redshift of 0.07. The results revealed no significant impact in the determination of the SFMS due to the extra consideration of the nebular emission. This result was attributed to the fact that, at low redshifts, normal SF galaxies (galaxies in the SFMS) do not experience a sufficiently high SF activity so that the nebular emission contributes significantly to the overall emission.

It thus seems that for galaxies hosting intense SF activity, accounting for the nebular emission is fundamental to accurately retrieve their physical properties. Conversely, for normal low-redshift SF galaxies, the nebular emission is not as relevant. Thus, it would be important to define a criterion to identify galaxies for which considering the nebular emission is indispensable. This is even of greater importance at increasingly higher redshifts, as a larger fraction of the galaxy population is expected to be experiencing a significant SF activity, when compared to galaxies in the local Universe.

The star formation history of the Universe shows that there was an epoch of rising SF activity that reached its maximum around z$\sim$1-3 and it has been steadily decreasing towards $z$=0 \citep{madau14}. Additionally, several works have studied the SFMS in a wide range of redshifts using different methods applied to data collected at various wavelengths (see \citealt{spea14} and \citealt{pop23} for a compilation of results). The results show that the slope of the SFMS does not evolve significantly with redshift. In contrast, the normalisation increases with redshift. This means that as the redshift increases, the general level of SF activity occurring in galaxies also increases.

Finally, the rest-frame EW of emission lines, which for SF galaxies is associated with the relative level of SF activity occurring within a galaxy, has also been shown to increase with redshift \citep[e.g.][]{fuma12,fais16,marmquer16,red18,sun23}. The increase in EW with redshift can become quite significant, reaching more than one order of magnitude at z$\sim$4. It is worth noting that a trend has also been observed between EW and stellar mass, with lower stellar mass galaxies having higher EWs \citep{salz89,kru95,fuma12}. This is a consequence of the fact that the ratio between the present and the average past star-formation rate (SFR) increases with decreasing stellar mass \citep{brinch04}. Hence, the EW value that is reached considering the redshift evolution depends on the stellar mass of the galaxy.

All these points suggest that at higher redshifts, the importance of modelling the nebular emission should be even greater, due to the higher level of SF activity occurring within galaxies. In fact, this warning has been highlighted in some works \citep[e.g.][]{scha09,scha10,atek11}. Furthermore, this can also be inferred from the evolution with redshift of the fraction of EELGs \citep[e.g.][]{boy24}. Hence, it is decisive to understand the role of nebular emission especially at higher redshifts, in order to identify for which galaxies modelling the nebular emission is fundamental and how the fraction of these objects evolves with redshift.

The importance of taking into consideration the nebular emission has also been the subject of interest due to the revolutionary observations of high-redshift galaxies by the \textit{James Webb} Space Telescope (JWST). Namely, studies have highlighted the substantial importance of incorporating nebular emission in templates used to select and characterise the physical properties of high-redshift galaxies \citep[e.g.][]{lar23,mcki23,wang24}. Also, JWST has enabled the census and characterisation of the physical properties of EELGs at $z$>3 \citep{dav23,boy24,ller24}. The results have shown highly SF systems, which may have had a crucial role in the reionisation of the Universe, with measured EWs beyond several hundred angstroms. Galaxies with such high EW measurements have also been observed in other studies that did not focus on EELGs \citep{cap23,robor24}.

JWST observations also revealed the existence of some elusive objects whose properties have been difficult to explain in light of our current knowledge. There have been several works addressing these objects and, in some cases, models with significant nebular emission have been used to explain their properties. For example, the recent discovery of the Ly$\alpha$-emitting galaxy JADES-GS+53.12175-27.79763 at z=5.943 \citep{sax24}, has sparked debate regarding the origin of its emission. \cite{cam24} proposes that the nebular lines and UV continuum emission of this galaxy can be explained by an initial mass function (IMF) that is between 10 to 30 times more top heavy than what is usually assumed. On the other hand, \cite{li24} proposes that a combination of a young and metal poor stellar population, plus a low-luminosity active galactic nucleus can also reproduce the observed emission. The different explanations lead to significantly different contributions of the nebular emission to the observed UV flux, with more than 80\% for the first case and 24\% for the latter case. JWST has also revealed a population of compact and extremely red objects, named 'little red dots' \citep{lab23,matt24}, whose nature  is still unclear. However, models with significant nebular continuum emission have been proposed to explain the peculiar features of these objects \citep[e.g.][]{iani24,pergon24}. All in all, the groundbreaking high-quality data from JWST is defying our understanding of galaxy and stellar population models and the role of the nebular emission is pivotal to explore these matters.

The significance of our study is manifold. Primarily, it will enable the identification of the population of galaxies for which the nebular emission plays a significant role at different redshifts. Moreover, some low-redshift EELGS, specifically extremely metal-poor BCDs, are local analogues of high-redshift galaxies. Therefore, accurately characterising their properties is fundamental to further comprehend the high-redshift Universe. Also, it is essential to understand how the fraction of galaxies for which the nebular emission is significant evolves with redshfit. Due to recent observation of numerous high-redshift galaxies by JWST, it has been pointed out that it is necessary to understand what are the best templates and procedures to model these objects. This work could also provide valuable insights to this question, specifically to the matter of the importance of taking into consideration the nebular emission when carrying out spectral fitting. These questions will become even more fundamental to address in light of future high-redshift surveys, such as MOONS \citep[Multi-Object Optical and Near-infrared Spectrograph;][]{moons1,moons2,moonrise}.

In this work, we  relate the contribution of the nebular to the total optical emission (hereafter referred to simply as the nebular contribution) with some commonly used tracers, such as the EW of Balmer lines, sum of emission line fluxes, SFR, and specific SFR (sSFR=SFR/M$_{*}$). Then, we study how the nebular contribution impacts the determination of physical properties of galaxies, specifically, the mass, age and stellar metallicity. Finally, we estimate the evolution with redshift of the fraction of galaxies for which we ought to take into account the nebular emission when carrying out spectral modelling.

The paper is organised as follows. Sect. \ref{sample} explains the sample selection. In Sect. \ref{method}, we describe the spectral fitting tools used to obtain the physical properties and nebular contribution estimates. Section \ref{Neb_cont} presents the estimation of the nebular contribution and its relationship with possible tracers and some physical properties. In Sect. \ref{results}, we study the impact of considering both stellar and nebular continuum on the physical properties of galaxies. In Sect. \ref{discussion}, we discuss the implications of these results. Lastly, in Sect. \ref{conclusion} the main results and implications of this work are summarised. We use a cosmology with H$_{0}$ = 70 km s$^{-1}$ Mpc$^{-1}$, $\Omega_{M}$=0.3, and $\Omega_{\Lambda}$=0.7, and we consider a \cite{chab03} IMF to estimate the star formation rate.

\section{Sample}
\label{sample}

To select our sample, we considered the results from the application of FADO to the data in SDSS-DR7 from \cite{card22}. This work provides measurements of emission line properties, plus estimates of physical and evolutionary properties for 926 246 galaxies from SDSS-DR7. The spectral fitting procedure used to obtain this data set is fully described in \cite{card22}.

To define our sample, we started by selecting galaxies that were properly fit with FADO by selecting the galaxies within the 0$<$$\chi^{2}$$<$1.5 interval. Then, to ensure that the ionising radiation is the product of star formation (and not of an AGN), we selected galaxies classified as SF following the BPT diagram \citep{bpt}. We also ensured that, for the selected galaxies, FADO detected the following emission lines: [OII] ($\lambda$ = 3727, 3729 Å), [OIII] ($\lambda$ = 4363, 4959, 5007 Å), H$\beta$, HeI ($\lambda$ = 5876 Å), [NII] ($\lambda$ = 6548, 6584 Å), H$\alpha$, and [SII] ($\lambda$ = 6717, 6731 Å). In the end, we obtained a sample of 173 457 galaxies, 19\% of the original sample.

Our aim was to obtain a sample that spans a wide range of values of both SF activity and nebular contribution. The EWs of Balmer lines are used in the literature as a tracer of both these quantities \citep[e.g.][]{papa98,gus07,atek11}. Hence, we considered the EW(H$\alpha$) as a tracer of the properties mentioned above to select our sample.

\begin{figure}[h!]
        \centering
        \includegraphics[scale=0.295]{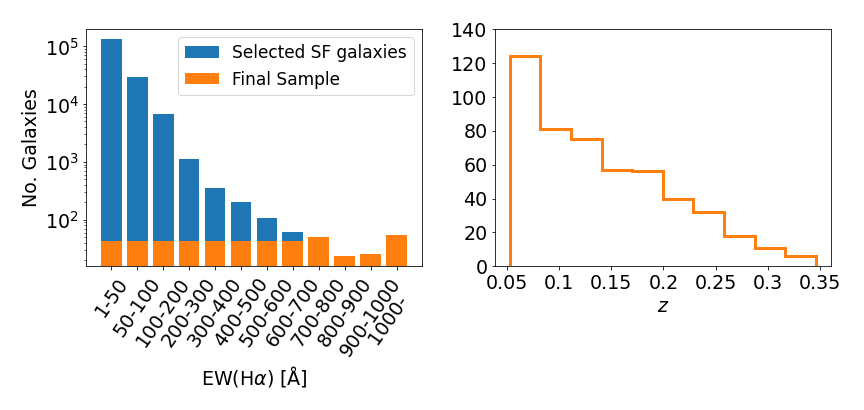}
        \caption{Characterisation of the samples considered in this work. \textit{Left panel}: EW(H$\alpha$) distribution for the selected SF galaxies, in blue, and for the final sample, in orange. \textit{Right panel}: Redshift distribution of the final sample.}
        \label{fig:Dist_Halpha}
    \end{figure}

We started by dividing our sample into bins of EW(H$\alpha$), as shown in the left panel of Fig. \ref{fig:Dist_Halpha} with the blue bars representing the distribution of our selected SF sample. The sample is highly skewed, with $\approx$95\% of the sample in the EW(H$\alpha$)$<$100 Å interval. However, for our study, we wanted a wide coverage of EW(H$\alpha$) values, from the lowest to the most extreme. Since for the highest EW(H$\alpha$) values there are not many galaxies in each bin, we included all of them in our final sample. On the other hand, for lower EW(H$\alpha$) bins, the number of galaxies is large; thus we randomly selected some of these objects in order to end up with a well-rounded sample of 500 galaxies evenly distributed in bins of EW(H$\alpha$). Figure \ref{fig:Dist_Halpha} shows, for our final sample, the EW(H$\alpha$) distribution (orange bars in the left panel) and the redshift distribution (right panel).

There are two important points regarding our final sample that are worth discussing: 1) it does not follow the overall distribution of the selected SF galaxies and 2) it has a small number of galaxies compared to the selected SF galaxies. Regarding the first point, this is consequence of the objective of this work. Given that we want a sample of galaxies that cover the widest possible range of SF activity and nebular contribution, there is no need to follow the distribution of the selected SF galaxies. In fact, if we had followed the overall distribution, we would have a final sample with a significant discrepancy between the number of galaxies with low and high EW(H$\alpha$) without changing the range covered by the sample. In relation to the second point, the size of our sample is limited by the number of galaxies available for the highest EW(H$\alpha$) bins. Additionally, since we are already selecting all of the available galaxies for the highest EW(H$\alpha$) bins, increasing the size of the sample would mean increasing the number of galaxies with low EW(H$\alpha$) values. As for the previous point, this would not increase the range covered by the sample.

\section{Method}
\label{method}

In general, the study of the importance of modelling the nebular emission when estimating the physical properties of galaxies has been put in terms of the level of SF activity of the galaxies. However, an absolute criterion required to characterise a galaxy as hosting an intense or low SF activity could be ambiguous. Firstly, because this is usually a relative characterisation which compares the properties of a galaxy relative to a reference value. For example, this characterisation can be done by comparing the position of a galaxy in the star formation rate-stellar mass diagram with the reference SFMS at that redshift \citep[e.g.][]{pop23}. Secondly, the reference point used to compare the properties of a galaxy may change with redshift. In the case of the SFMS, it is well established the increase in the normalisation with increasing redshift. In this way, defining the necessity of modelling the nebular emission in terms of the nebular contribution could be a better approach.

We  used the spectral fitting tool FADO to fit the spectra of the galaxies in our sample. FADO, with its self-consistent approach to the fitting of the optical spectrum of galaxies, provides a way to obtain a stellar and nebular model that are consistent with each other. FADO is capable of ensuring this self-consistent approach to spectral fitting through the combination of its coherent modelling of the most important nebular features, such as the H$\alpha$ and H$\beta$ luminosities and EWs and the continuum around the Balmer and Paschen jumps. The nebular spectrum itself is computed using standard photoionisation prescriptions \citep[e.g.][]{kru95} considering the contribution of two-photon, free-free and free-bound emission \citep[see][for further details]{FADO}. FADO also offers the possibility of being used in pure stellar mode, meaning it can be invoked such that the fit is done assuming only a stellar continuum. The mathematical concept of FADO, in particular the search for the best-fitting solution using differential evolution optimisation \citep{sto96}, is a further asset. Thus, FADO is an appropriate tool to obtain a model of the stellar and nebular continuum of the galaxies in our sample in order to estimate the fraction of the nebular emission.

With this work, we aim to study how considering the nebular emission from galaxies impacts the estimation of their physical properties. To do so, we also need a spectral fitting code that lacks the self-consistency concept of FADO (i.e. a code driven solely by $\chi^2$ minimization between model and observed continuum) and invoked such as to only take stellar emission into account. We used STARLIGHT\footnote{We note that STARLIGHT can be invoked with a library that additionally contains non-stellar spectral templates, such as a nebular continuum component \citep{corb06} or a power law \citep[e.g.][]{card16,riff22}.} \citep{cid05}, as representative of this type of codes. In our application, STARLIGHT only considers a linear combination of simple stellar populations (SSPs) to create a model spectrum which is fitted to the observed spectrum using a minimisation process carried out through the combination of simulated annealing with the Metropolis and Markov chain Monte Carlo (MCMC) techniques. These two codes have been compared in the past with the aim of understanding the role of nebular emission in spectral fitting \citep{FADO,card19,card22,breda22}, thereby supporting our methodology.

Before the spectral fitting, we corrected the spectra for Galactic extinction using the dust maps from \cite{schle98}, plus the correction factor from \cite{schla11}, and we considered the \cite{carde89} extinction curve. Then, we converted the spectra to the rest-frame and rebinned it such that $\Delta \lambda$ = 1 \AA. This rebinning was performed since it is recommended for STARLIGHT. Although not required for FADO, the spectra needed to be resampled so that both codes could be applied consistently.

For the application of FADO and STARLIGHT, we considered the $\lambda$=3000--9000 Å spectral range with a spectral basis composed of 171 SSPs, derived from \cite{bruz03} considering a \cite{chab03} IMF and Padova 1994 evolutionary tracks \citep{alo93,bres93,faga94,fagb94,gir96}, and the \cite{calz00} extinction law for dust attenuation corrections. The spectral basis contains 57 ages (t = [0.5 Myr, 13 Gyr]) and 3 metallicities (Z = 0.2, 0.4, 1 Z$_{\odot}$). Additionally, we allowed in the fitting the V-band extinction to vary between 0 and 4 magnitudes. Both the spectral basis and the V-band extinction interval are different relatively to the spectral fitting parameters used in \cite{card22}. Regarding the masking of the most prominent emission lines, FADO performs this process automatically as part of its internal procedure; whereas for STARLIGHT, these spectral regions were manually identified and provided as input for the fitting. The masked regions are consistent between both codes.

Regarding the selected spectral basis, we chose a set of reasonable SSPs for the modelling of galaxies with both low and significant nebular emission. Moreover, since young SSPs contribute more significantly to the overall nebular emission, we considered all the SSPs available in \cite{bruz03} in the critical period between 0.5--15 Myr. To ensure that our conclusions do not depend on the chosen spectral basis, we repeated the analysis using a spectral basis with 150 SSPs containing 25 ages (t = [1 Myr, 15 Gyr]) and 6 metallicities (Z = 0.005, 0.02, 0.2, 0.4, 1, 2.5 Z$_{\odot}$). The obtained results are comparable using both spectral bases and we can confirm that the results derived in this work are independent of the chosen spectral basis.

With the aim of obtaining an adequate error estimation for the parameters considered in this work, we perturbed the observed spectrum and repeated the FADO run 50 times for each galaxy. To perturb the spectrum, we considered that the error associated with each flux measurement followed a Gaussian distribution centred around zero with standard deviation equal to the error in the flux measurement. After perturbing the spectra and repeating the FADO run, we consider the error as the standard deviation of the values obtained for the runs performed for each galaxy.

As products of the application of FADO, we will use the continuum models for the total, stellar and nebular components. We will also use the flux and EW measurements for several emission lines and the estimated physical properties of galaxies, specifically the currently available and total ever formed stellar mass (M$_{curr}$ and M$_{ever}$, respectively), light- and mass-weighted stellar ages (t$_{L}$ and t$_{M}$, respectively) and light- and mass-weighted stellar metallicities (Z$_{L}$ and Z$_{M}$, respectively). From the application of STARLIGHT, we will use the aforementioned physical properties of galaxies.

One important aspect to take into consideration is the fact that SDSS is a fibre-based survey. This means that in some cases the fibre aperture (3 arcsec) might not be sufficient to completely cover the galaxy. Since the aim of this work is to understand the impact of considering the nebular emission in spectral modelling, and not to accurately estimate the physical properties of galaxies as a whole, we do not correct for aperture effects. Additionally, even if we corrected the spectra for aperture effects, in the end we would be applying FADO and STARLIGHT to the same spectrum and comparing the different estimates obtained by the two codes. Thus, the results presented in this work do not depend on aperture effects. Henceforth, when we mention galaxy properties we are referring to the properties of the galaxy region covered by the fibre.

Another aspect that is worth highlighting is that the EW measurements of FADO are obtained following a different continuum definition in relation to what is commonly used in the literature. FADO measures the EW considering that the continuum is the monochromatic continuum level at the central wavelength of the line, thus accounting for possible absorption features and stellar velocity dispersion. However, many works in the literature measure the EW based on a continuum that is estimated considering the average level of the continuum adjacent to the considered spectral line, which is called the pseudo-continuum. In \cite{mir23}, these two approaches to estimate the EW were compared using FADO and data from the MPA-JHU catalogue. It was found that the approach followed in FADO to measure EWs leads to estimates that are on average 0.1 dex ($\sim$25\%) higher relative to the other approach. Hence, EW measurements obtained based on the pseudo-continuum level need to be scaled by this factor to be comparable with the relations and results derived in this work, since we use the EW estimates from FADO.

\section{Nebular contribution, its tracers, and the physical properties of galaxies}
\label{Neb_cont}

\begin{figure}[h!]
        \centering
        \includegraphics[scale=0.35]{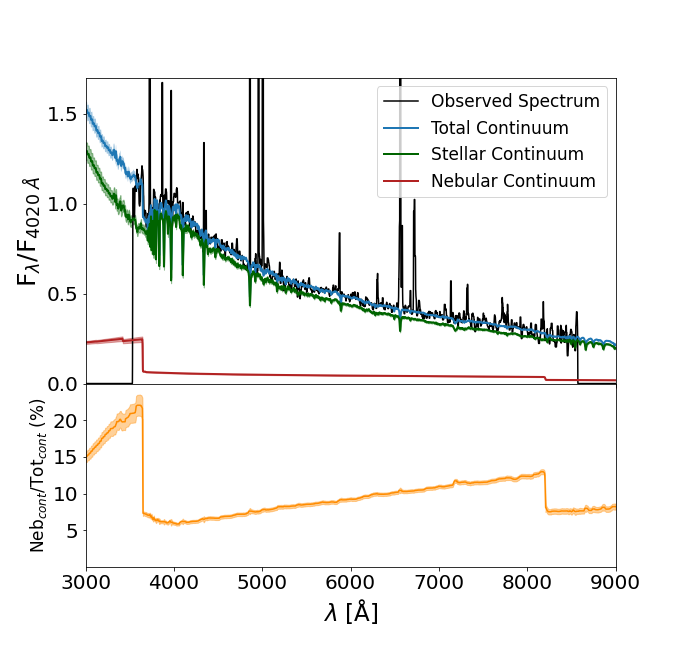}
        \caption{Model fitted by FADO to a galaxy in our sample with X$_{neb}$=10\%. \textit{Upper Panel:} Observed spectrum (black line) and total, stellar and nebular continuum fitted by FADO (blue, green and red lines, respectively). The shaded regions represent the uncertainty in the estimated models. \textit{Bottom Panel:} Ratio between the nebular and total continuum fitted by FADO as a function of wavelength.}
        \label{fig:Xneb_vs_lambda}
    \end{figure}

The upper panel of Fig. \ref{fig:Xneb_vs_lambda} shows the observed spectrum (black line), and the total, stellar and nebular continuum (blue, green and red lines, respectively) fitted by FADO, for one of the galaxies in our sample. In the bottom panel, the nebular contribution as a function of wavelength is plotted for that same galaxy.

The definition of median nebular contribution (X$_{neb}$) used throughout this work is obtained by computing the ratio between the nebular and total continuum at each wavelength, and then calculating the median value of this ratio in the 3000-9000 \AA\ interval. It is worth considering that the contribution of the nebular to the total continuum changes significantly with wavelength (see the bottom panel of Fig. \ref{fig:Xneb_vs_lambda} and also \citealt{izo11}). This is particularly noticeable in the region before and after the Balmer and Paschen discontinuities, with a sudden and conspicuous drop in the contribution. Nonetheless, considering a median level of nebular contribution provides a way to have an estimate of the importance of the nebular emission to the total observed emission of a galaxy in the optical regime.

\begin{figure}[h!]
        \centering
        \includegraphics[scale=0.3]{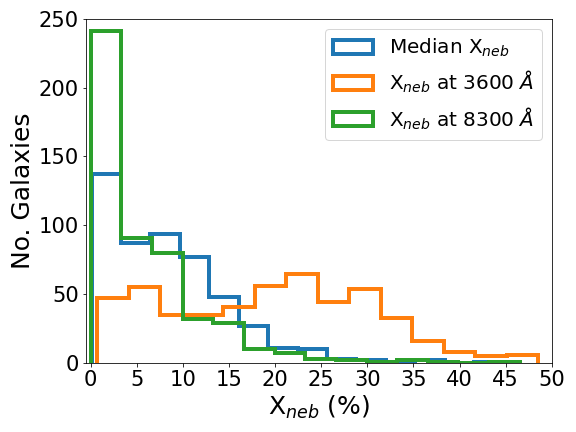}
        \caption{Distribution of the estimated median nebular contribution (blue) and at 3600 \AA\ and 8300 \AA\ (orange and green, respectively) for the galaxies in our sample.}
        \label{fig:Xneb_dist}
    \end{figure}

In Fig. \ref{fig:Xneb_dist}, we show the X$_{neb}$ distribution for the galaxies in our sample. We also show the nebular contribution at 3600 \AA, blueward of the Balmer discontinuity, and at 8300 \AA, after the Paschen discontinuity. These two wavelengths sample the regions where the nebular contribution in the optical spectral range is close to its maximum and minimum, respectively. As expected, the distribution of the nebular contribution at 3600 \AA\ extends towards higher values relative to the median X$_{neb}$, whereas the nebular contribution at 8300 \AA\ has more occurrences around lower values.

The galaxies in our sample have an average nebular contribution of X$_{neb}$=8\%. The distribution shows that we have an adequate coverage of the parameter space, even if many galaxies (38\%) have X$_{neb}$$<$5\% and the number of galaxies drops dramatically as X$_{neb}$ increases, with only 14\% having X$_{neb}$$>$15\%.

\subsection{Tracers of the nebular contribution}

The analysis carried out in this work is highly demanding in terms of quality of the spectrum and continuum detection. This is not always possible, particularly at higher redshifts. Hence, it is important to understand which nebular contribution tracers, among the most commonly used in the literature, are more adequate, and to quantify this relation. This will enable the possibility of having an estimate of the optical nebular contribution simply by using a tracer.

We considered nebular contribution tracers related with emission line properties, namely the EW of H$\alpha$ and H$\beta$, considering for each case only the galaxies where these lines had S/N$>$3 (for H$\alpha$ all galaxies meet the criteria whereas for H$\beta$, 480 galaxies are selected). We also considered the sum of the flux of the most prominent emission lines in the optical \citep{breda22}, specifically: [OII] ($\lambda$ = 3727, 3729 Å), [OIII] ($\lambda$ = 4363, 4959, 5007 Å), H$\beta$, HeI ($\lambda$ = 5876 Å), [NII] ($\lambda$ = 6548, 6584 Å), H$\alpha$, and [SII] ($\lambda$ = 6717, 6731 Å). In this case, we did not apply the S/N$>$3 restriction since some of these lines are weak and the sample would be significantly reduced. Finally, we also considered as tracers the current SFR and sSFR.

We started by correcting the fluxes of all emission lines for the intrinsic extinction using the Balmer decrement (H$\alpha$/H$\beta$) and following the procedure described in \cite{mir23}. To estimate the current SFR, we used the H$\alpha$ flux as a tracer. First, we calculated the H$\alpha$ luminosity and then, to obtain the SFR, divided it by the \cite{ken98} conversion factor, calibrated for a \cite{chab03} IMF: $\eta$(H$\alpha$)=10$^{41.31}$ erg s$^{-1}$ M$_{\odot}^{-1}$ yr \citep{mir23}. There are other conversion factors that could have been used \citep[e.g][]{wilk19}. However, using a different calibration would change the SFR estimate for all galaxies by the same amount and either to lower or higher values. Thus, this would only impact the relations derived in this work in term of normalisation, but not in terms of the observed scatter. The sSFR is obtained by computing the ratio between the estimated SFR and the currently available stellar mass estimated by FADO.

It is noteworthy that the \cite{ken98} conversion factor between SFR and H$\alpha$ luminosity assumes a continuous SFR over more than 80 Myr and that the metallicity of the galaxy is similar to the solar metallicity. For galaxies with sub-solar metallicities which deviate substantially from these assumptions, such as EELGs, this leads to an overestimation of the SFR \citep{weil01,papa23}. Although many of the galaxies in our sample have sub-solar metallicities, they are not extremely metal-poor. Thus, even though our SFR estimates are likely affected by errors, these should not be substantial enough to significantly affect the relationships established in this study.

\begin{figure*}[h!]
        \centering
        \includegraphics[scale=0.38]{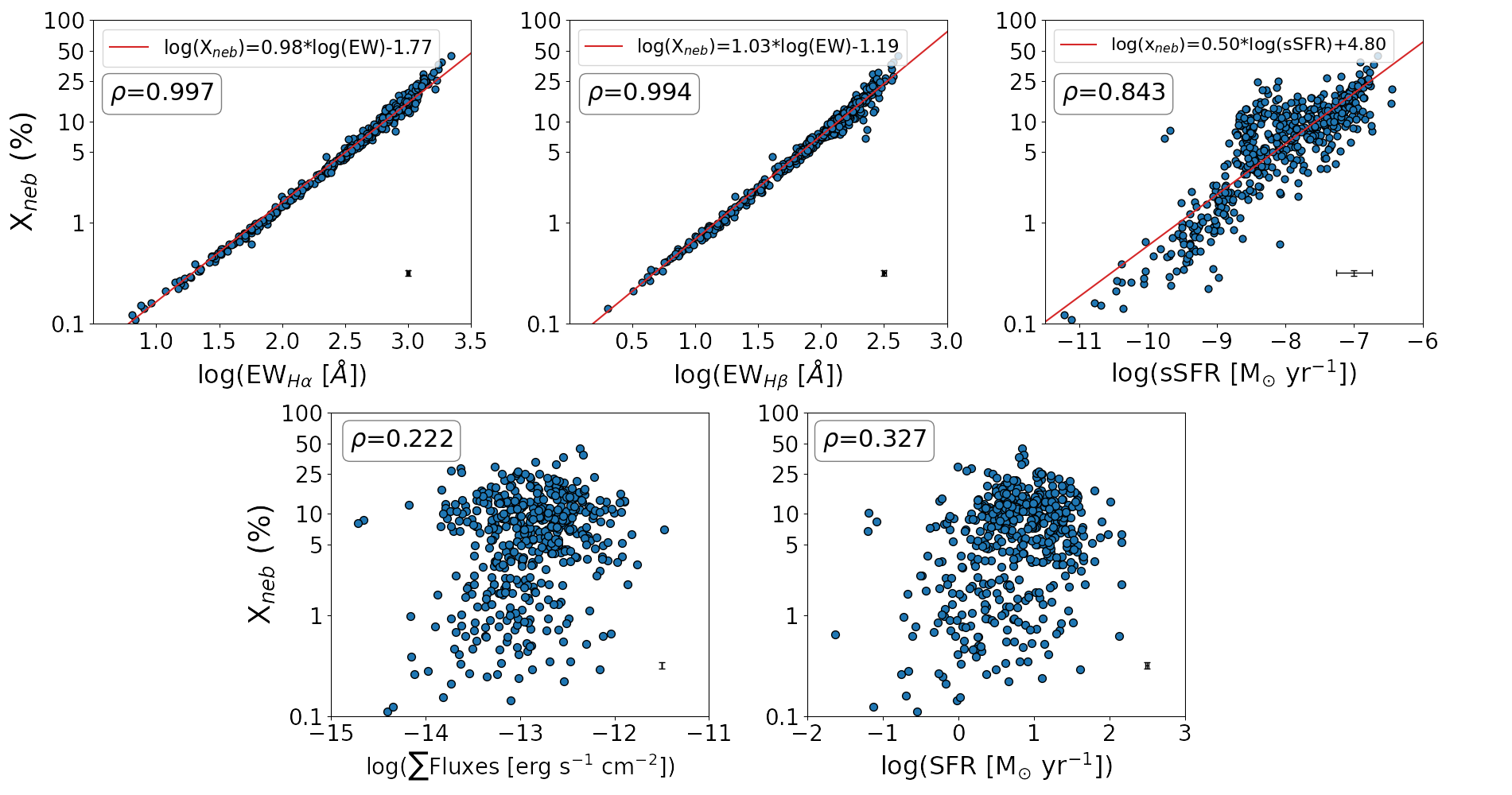}
        \caption{Relation between X$_{neb}$ and the different tracers considered in this work: EW of H$\alpha$ (upper left), EW of H$\beta$ (upper middle), sSFR (upper right), sum of the flux of the most prominent emission lines (lower left) and SFR (lower right). In the upper left of each panel, we present the Pearson correlation coefficient assuming a linear fit to the data. For the selected best tracers of X$_{neb}$, the red line is the derived best linear fit and the resulting equation is presented in the legend. The median error of the data is presented in the bottom right corner. The axes are in logarithmic scale.}
        \label{fig:Xneb_vs_tracers_fit}
    \end{figure*}

In Fig. \ref{fig:Xneb_vs_tracers_fit}, X$_{neb}$ is plotted as a function of different potential nebular contribution tracers. For all tracers studied, there is a positive correlation with X$_{neb}$ even if for the sum of fluxes and SFR, the scatter is significant. Indeed, the Pearson correlation coefficient for these two tracers further points to a weak correlation with X$_{neb}$.

Among the selected tracers of the nebular contribution, the EW of H$\alpha$ and H$\beta$ clearly stand out from the others, having a strong linear correlation and minor scatter. The EW is the ratio between the line flux and the continuum level. For SF galaxies, this means that it is also a measure of the relation between the current SFR, more closely represented by the line flux, and the average SFR throughout the lifetime of the galaxy, approximately represented by the continuum level. Therefore, this result is understandable and in line with the literature \citep{papa98,gus07,atek11}.

Finally, the sSFR also shows a clear correlation with the nebular contribution, even though the scatter is larger than the one for the EW of H$\alpha$ and H$\beta$. Since there is a significant scatter for the SFR, but not so large for the sSFR, this means that the stellar mass contributes significantly to the observed scatter. Galaxies of significantly different stellar masses can have the same SFR, hence the nebular emission associated with the SF activity will also be the same. However, the contribution of the nebular emission to the total emission will depend on the stellar mass, since the stellar mass is connected to the level of the stellar emission. By combining the SFR with the stellar mass, this factor is removed and consequently the scatter becomes lower.

The clear correlation between the sSFR and X$_{neb}$ suggests that the former can be used as a tracer of the latter. However, there are some disadvantages associated. First, our sSFR estimates were obtained from the H$\alpha$ luminosity, and might not be directly comparable with sSFR estimates obtained from other methods. Also, whereas the EW of H$\alpha$ and H$\beta$ can be directly measured, the sSFR depends on the method used to estimate both SFR and stellar mass, thus increasing the uncertainties associated with its estimate. Consequently, the sSFR is not the ideal tracer of the nebular contribution. Nevertheless, it can still be useful to provide an indication of the approximate level of the nebular contribution. For the specific case of galaxies that only have photometric data, and thus might not be possible to estimate either the H$\alpha$ or H$\beta$ EW, their sSFR can be estimated and used to have an approximate estimate of the level of nebular contribution.

In summary, from the analysed nebular contribution tracers, our results show that the EW of H$\alpha$ and H$\beta$ are the most appropriate, whereas the sSFR emerges as a possible alternative, with the previously mentioned caveats. Given the observed linear correlation between X$_{neb}$ and these tracers, we applied a linear fit to the data and the results are presented in the upper panels of Fig. \ref{fig:Xneb_vs_tracers_fit}. The best-fitting equations obtained are the following:

\vspace{-0.4cm}
    \begin{equation}
        \log(\text{X}_{neb} [\%]) = 0.98\pm0.05 \times \log(\text{EW}_{\text{H}\alpha} [\AA]) - 1.77\pm0.13,
        \label{eq:Xneb_Ha}
     \end{equation}
\vspace{-0.6cm}
    \begin{equation}
        \log(\text{X}_{neb} [\%]) = 1.03\pm0.03 \times \log(\text{EW}_{\text{H}\beta} [\AA]) - 1.19\pm0.05,
        \label{eq:Xneb_Hb}
    \end{equation}
\vspace{-0.6cm}
    \begin{equation}
        \log(\text{X}_{neb} [\%]) = 0.50\pm0.03 \times \log(\text{sSFR} [yr^{-1}]) + 4.80\pm0.24
        \label{eq:Xneb_sSFR}
    .\end{equation}

\noindent These equations allow for the estimation of X$_{neb}$ without the need for carrying out spectral fitting. This is particularly useful for high-redshift galaxies for which (in most cases) the continuum has not been detected and the nebular emission might play a significant role.

\begin{figure}[h!]
        \centering
        \includegraphics[scale=0.3]{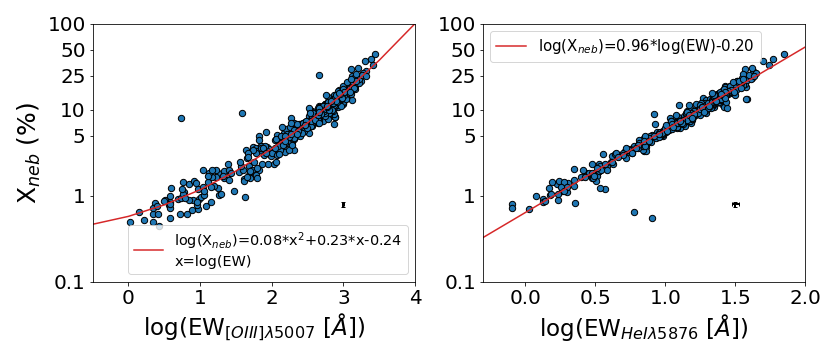}
        \caption{Relation between X$_{neb}$ and EW of [OIII]$\lambda5007$ (left panel) and HeI$\lambda$5876 (right panel) in log-log scale. The red line is the derived best fit to the data and the resulting equation is presented in the legend. The median error of the data is presented in the bottom right corner.}
        \label{fig:Xneb_vs_tracers2_fit}
    \end{figure}

In addition to the EW of H$\alpha$ and H$\beta$, we also plotted X$_{neb}$ as a function of the EW of other optical emission lines. Among the considered ones, [OIII]$\lambda$5007 and HeI$\lambda$5876 stand out as the emission lines with EWs that are best correlated with X$_{neb}$, probably due to being related with young massive stars ionising the gas. The [OIII]$\lambda$5007 emission line is commonly used to select EELGs and the HeI$\lambda$5876 emission line is relatively strong in low-metallicity SF galaxies, thus it is interesting that these lines show a strong relation with X$_{neb}$. Before fitting the relation, we applied the same S/N$>$3 criteria as before, selecting 466 and 397 galaxies for [OIII]$\lambda$5007 and HeI$\lambda$5876, respectively. In Fig. \ref{fig:Xneb_vs_tracers2_fit}, we show these correlations and the resulting fits.

The relation between X$_{neb}$ and the EW([OIII]$\lambda$5007) is not linear, being better described by a second-order polynomial. This suggests that the connection between the EW([OIII]$\lambda$5007) and the EW of H$\alpha$ and H$\beta$ also follows a second-order polynomial. Possibly, this is related with the dependence on metallicity of the [OIII]$\lambda$5007 emission line. However, this analysis is beyond the scope of the present paper and should be continued in a future work. In the case of the EW(HeI$\lambda$5876), there is a linear relation with X$_{neb}$. The best-fitting equations obtained are as follows:

    \vspace{-0.4cm}
    \begin{equation}
        \begin{aligned}
            \log(&\text{X}_{neb} [\%]) = 0.08\pm0.01 \times \log(\text{EW}_{\text{[OIII]}\lambda5007} [\AA])^{2} + \\
            &+ 0.23\pm0.04 \times \log(\text{EW}_{\text{[OIII]}\lambda5007} [\AA]) - 0.24\pm0.04, \\
        \end{aligned}
    \end{equation}
\vspace{-0.2cm}
    \begin{equation}
        \log(\text{X}_{neb} [\%]) = 0.96\pm0.02 \times \log(\text{EW}_{\text{He}\lambda5876} [\AA]) - 0.20\pm0.03.
    \end{equation}

\subsection{Physical properties of galaxies and the nebular contribution}

It is also interesting and relevant to understand how the nebular contribution relates to different physical properties of galaxies, namely the stellar mass, age, and metallicity, along with the gaseous metallicity. We used the currently available stellar mass and light-weighted age and metallicity estimates from FADO. We estimate the gaseous metallicity, Z$_{gas}$, by calculating the tracer R$_{23}$=([OII]$\lambda$3727+[OIII]$\lambda$4959, 5007)/H$\beta$, using the flux measurements from FADO, and following the relation defined in \cite{naka22}.

\begin{figure}[h!]
        \centering
        \includegraphics[scale=0.27]{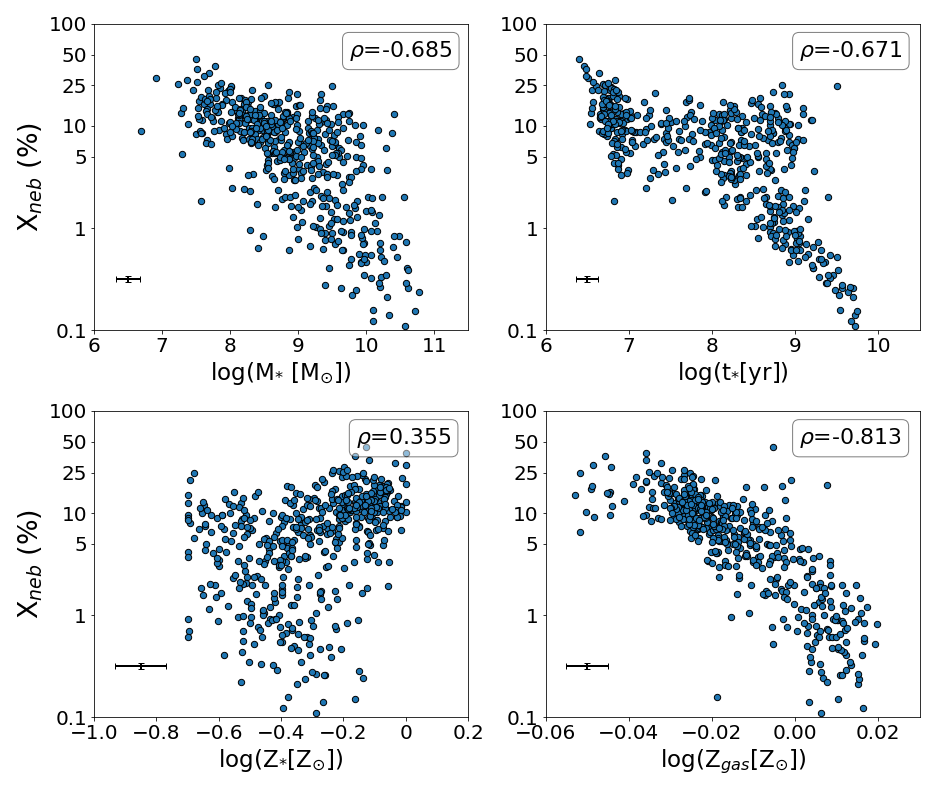}
        \caption{Relation between X$_{neb}$ and currently available stellar mass (upper left), light-weighted stellar age (upper right), light-weighted stellar metallicity (bottom left) and gaseous metallicity (bottom right). In the upper right of each panel, we present the Pearson correlation coefficient assuming a linear fit to the data. The median error of the data is presented in the bottom left corner.}
        \label{fig:Xneb_vs_gal_prop}
    \end{figure}

In Fig. \ref{fig:Xneb_vs_gal_prop}, the relation between X$_{neb}$ and the currently available stellar mass, light-weighted age and metallicity and gaseous metallicity is plotted. The stellar mass and age and gaseous metallicity show a negative correlation with X$_{neb}$. This shows that, as expected, high nebular contributions are particularly associated with young, low-mass and metal-poor galaxies. Moreover, considering the obtained relations between the nebular contribution and sSFR, stellar mass and gaseous metallicity we observe the same "downsizing" evidence reported in \cite{mannu10}, where high stellar mass galaxies have lower sSFR and higher gaseous metallicities.

Our analysis does not reveal a clear correlation between the stellar metallicity and X$_{neb}$, which is also indicated by the low value of the Pearson correlation coefficient. For the relation between the gaseous metallicity and X$_{neb}$, the Pearson correlation coefficient points towards a rather moderate to strong correlation. However, the scatter is still large enough so that it is not an ideal tracer of X$_{neb}$. In the end, we can hypothesise that the gaseous metallicity could be used to obtain a broad indication of the nebular contribution level. 

\section{Nebular component impact threshold}
\label{results}

One of the aims of this work is to derive a threshold above which the contribution of the nebular to the total emission becomes significant, and considering it leads to non-negligible differences in the derived physical properties of galaxies. For this analysis, we calculated the difference between the estimates of FADO and STARLIGHT for the different physical properties. Then, we divided our sample according to the EW(H$\alpha$) estimated by FADO, which in Sect. \ref{Neb_cont} we have shown to be a strong tracer of the nebular contribution, in the following four bins: EW(H$\alpha$)$<$100 \AA, 100$\leq$EW(H$\alpha$)$<$500 \AA, 500$\leq$EW(H$\alpha$)$<$1000 \AA\ and EW(H$\alpha$)$\geq$1000 \AA.

Figures \ref{fig:FD_vs_ST_M}, \ref{fig:FD_vs_ST_t}, and \ref{fig:FD_vs_ST_Z} show the distribution of the logarithm of the ratio between FADO (FD) and STARLIGHT (ST) estimates of different physical properties of galaxies: stellar mass, age and metallicity, respectively. The distributions are normalised such that the area under the curve is equal to one, thus allowing to directly compare the distributions of the different EW(H$\alpha$) bins. The distributions will help identify when FADO obtains larger estimates relative to STARLIGHT (positive side of the plots) and the opposite case (negative side of the plots). Additionally, we show as a vertical dashed line the median value of each distribution.

Table \ref{tab:FD_vs_ST} shows the percentage of objects for which either FADO or STARLIGHT obtains an estimate that is significantly different relative to the estimate of the other code. This will complement the information given by the previously mentioned distributions and give further insights into how the distributions change with varying level of nebular contribution. FADO and STARLIGHT can retrieve the studied physical properties within a typical accuracy of 0.2 dex. This value is an upper limit of the tested accuracy of both codes when retrieving the analysed physical properties of galaxies \citep{FADO,card19,ciro21,cid05,cid14}. Hence, we consider that FADO and STARLIGHT obtain significantly different estimates when they are larger than 0.2 dex.

\subsection{Stellar mass}

\begin{figure}[h!]
        \centering
        \includegraphics[scale=0.35]{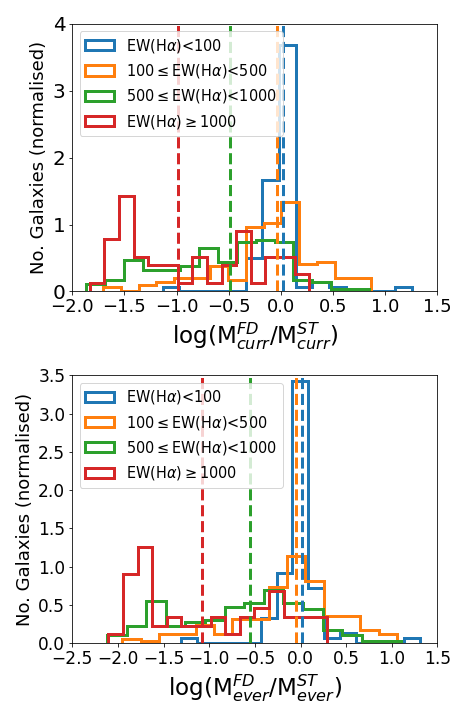}
        \caption{Distribution of the logarithmic difference between the currently available stellar mass (upper panel) and total ever formed stellar mass (bottom panel) estimated by FADO (FD) and STARLIGHT (ST), for four EW(H$\alpha$) bins: EW(H$\alpha$)$<$100 \AA\ (blue), 100$\leq$EW(H$\alpha$)$<$500 \AA\ (orange), 500$\leq$EW(H$\alpha$)$<$1000 \AA\ (green) and EW(H$\alpha$)$\geq$1000 \AA\ (red). The vertical dashed lines represent the median value of each distribution.}
        \label{fig:FD_vs_ST_M}
    \end{figure}

Figure \ref{fig:FD_vs_ST_M} shows the evolution with the level of nebular contribution of the distribution of the difference between FADO and STARLIGHT for the currently available and total ever formed stellar mass. The distributions are similar for both quantities and evolve in a comparable manner and can be analysed together.

For the first interval, EW(H$\alpha$)$<$100 \AA, the distributions are centred around zero, with less than 15\% of galaxies being significantly different between the two codes. As we move to the higher EW(H$\alpha$) intervals, the distribution becomes broader, with an increasing number of galaxies being significantly different between the two codes, and the peak around zero becomes less prominent. However, the distribution does not broaden uniformly, it moves towards the negative side of the plot, which represents STARLIGHT obtaining larger estimates than FADO. Although for the 100$\leq$EW(H$\alpha$)$<$500 \AA\ interval there is an increase in the percentage of galaxies significantly different between the two codes in both directions, even if with a slight tendency for STARLIGHT obtaining higher estimates than FADO, in the two intervals of higher EW(H$\alpha$) a trend is clearly established. In the 500$\leq$EW(H$\alpha$)$<$1000 \AA\ and EW(H$\alpha$)$\geq$1000 \AA\ intervals, STARLIGHT estimates a higher stellar mass for the vast majority of galaxies (more than 70\%), while for only a few galaxies does it estimate lower stellar masses (less than 7\%). This is evident in the distributions, particularly with the appearance of a peak in the region of STARLIGHT estimating higher values, around $\log(M^{FD}/M^{ST})$=-1.5 dex.

The results show that for EW(H$\alpha$)$\geq$500 \AA, there is a clear trend for STARLIGHT to obtain higher stellar mass estimates relative to FADO, with a mean difference of $\sim$0.6 dex. Considering only the most extreme cases, EW(H$\alpha$)$\geq$1000 \AA, then the mean difference is $\sim$1 dex and can reach values up to 2 dex. Thus, we can identify the value EW(H$\alpha$)=500 \AA\ as the threshold for which a statistically significant difference between the two codes is identifiable.

\subsection{Stellar age}

\begin{figure}[h!]
        \centering
        \includegraphics[scale=0.35]{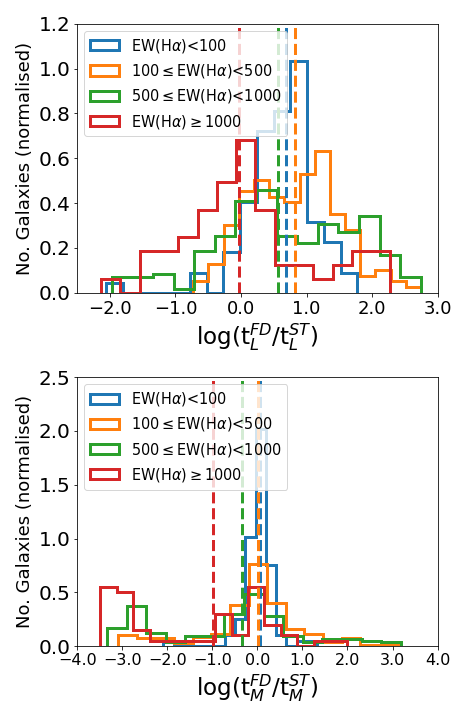}
        \caption{Same as Fig. \ref{fig:FD_vs_ST_M}, but for the light- and mass-weighted stellar age (upper and lower panels, respectively).}
        \label{fig:FD_vs_ST_t}
    \end{figure}

Figure \ref{fig:FD_vs_ST_t} shows the evolution with the level of nebular contribution of the distribution of the difference between FADO and STARLIGHT for the light- and mass-weighted stellar age. In this case, the distribution of these quantities evolves differently, therefore we will analyse them separately.

Considering the light-weighted stellar age, in the first three EW(H$\alpha$) bins, FADO estimates higher values relative to STARLIGHT for most galaxies. On the other hand, for the EW(H$\alpha$)$\geq$1000 \AA\ interval, the percentage of objects where one of the codes obtains significantly higher estimates relative to the other is comparable. Looking at the distributions, it seems that except for the EW(H$\alpha$)$<$100 \AA\ interval, they become increasingly bimodal with increasing EW(H$\alpha$), with one of the peaks moving towards higher values and the other towards lower values. For the highest EW(H$\alpha$) bin, one of the peaks is around zero while the other is at $\log(t_{L}^{FD}/t_{L}^{ST})$$\simeq$2 dex. This suggests that as the EW(H$\alpha$) increases, there is a population of galaxies for which FADO and STARLIGHT estimates become compatible, while for another population FADO obtains increasingly larger estimates relative to STARLIGHT.

Considering the mass-weighted stellar age, for the first two EW(H$\alpha$) bins there is a similar percentage of objects which either one of the codes obtains significantly larger estimates relative to the other. For the next two EW(H$\alpha$) bins, STARLIGHT obtains larger estimates than FADO for at least 54\% of the objects. Analysing the distributions for further insights, it is seen that for all EW(H$\alpha$) bins there is a peak around zero, showing that for a subsample of galaxies both codes obtain compatible estimates. However, as the EW(H$\alpha$) increases, a peak arises and becomes increasingly more pronounced in the region of the distribution corresponding to FADO obtaining lower estimates than STARLIGHT, with differences larger than 2 dex. This feature is also observed in Fig. \ref{fig:FD_vs_ST_M} for the EW(H$\alpha$)$\geq$1000 \AA\ interval.

In summary, regarding the light-weighted stellar age, for galaxies with EW(H$\alpha$)$<$1000 \AA\ there is a trend for FADO to obtain larger estimates compared to STARLIGHT, with a mean difference of $\sim$0.7 dex. Contrarily, for mass-weighted stellar ages, the trend is for STARLIGHT to estimate larger values when EW(H$\alpha$)$\geq$500 \AA, obtaining a mean difference of $\sim$0.9 dex. However, the distribution becomes bimodal, with FADO estimating significantly higher values for a subset of galaxies. Therefore, these limits represent thresholds for a significant difference between the estimates of the two codes.

\subsection{Stellar metallicity}

\begin{figure}[h!]
        \centering
        \includegraphics[scale=0.35]{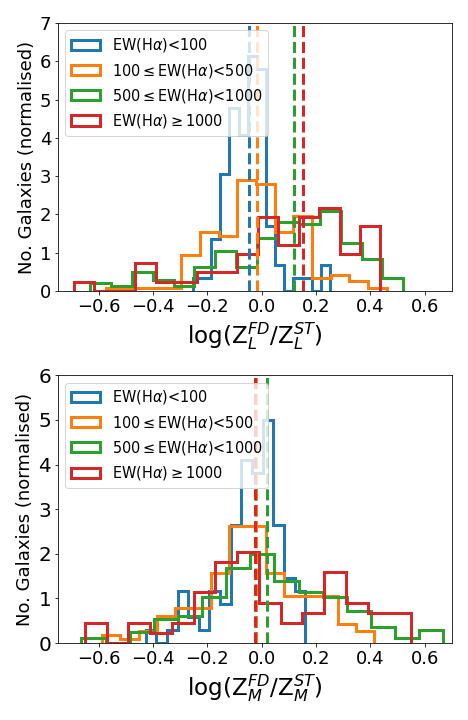}
        \caption{Same as Fig. \ref{fig:FD_vs_ST_M}, but for the light- and mass-weighted stellar metallicity (upper and lower panels, respectively).}
        \label{fig:FD_vs_ST_Z}
    \end{figure}

Figure \ref{fig:FD_vs_ST_Z} shows the evolution with the level of nebular contribution of the distribution of the difference between FADO and STARLIGHT for the light- and mass-weighted stellar metallicity. Both quantities have close distributions and have an analogous evolution, so we will analyse them together.

For all defined EW(H$\alpha$) intervals, the percentage of galaxies for which one of the codes estimates significantly different metallicities relative to the other only reaches 50\%. Nonetheless, as the EW(H$\alpha$) increases, there is a slight tendency for FADO to obtain higher metallicity estimates. In fact, for EW(H$\alpha$)$\geq$500 \AA\ FADO obtains larger estimates for around 30\% of the sample. This fact can also be seen in the distributions, which tend to shift to the part of the plot corresponding to FADO estimating higher values than STARLIGHT, and in the mean value evolution. However, a peak around zero remains in all distributions, highlighting that FADO and STARLIGHT measurements are compatible for a significant proportion of the galaxies. To further emphasise this point, we find that the mean differences between the two codes are 0.03 and 0.003 dex for the light- and mass-weighted metallicity, respectively. Even for the EW(H$\alpha$)$\geq$1000 \AA\ interval, the mean differences only reaches 0.09 dex, well within the defined uncertainty level.

Summarising, there is a trend for FADO obtaining higher metallicities relative to STARLIGHT. However, it does not reach the point where FADO estimates higher values for most of the galaxies or that the mean differences are larger than the uncertainty level, even when considering the EW(H$\alpha$)$\geq$1000 \AA\ interval. Hence, there is no clear threshold above which we observe a significant difference between the estimates of the two codes.

    \begin{table*}[h!]
        \centering
        \caption{Analysis of the amount of galaxies for which there is significant difference between the FADO and STARLIGHT estimates of the currently and total ever-formed stellar mass, light- and mass-weighted stellar age, and metallicity.}
        \begin{tabular}{cc|c|c|c|c}
            \multicolumn{2}{c|}{}
            & EW$_{H\alpha}$$<$100 & 100$\leq$EW$_{H\alpha}$$<$500 & 500$\leq$EW$_{H\alpha}$$<$1000 & EW$_{H\alpha}$$\geq$1000 \\ \hline
            \multirow{2}{*}{M$_{curr}$}  & FD & 5\%  & 21\%  & 5\%  & 2\%  \\
                                         & ST & 9\%  & 34\%  & 71\% & 82\% \\ \hline
            \multirow{2}{*}{M$_{ever}$}  & FD & 5\%  & 22\%  & 7\%  & 4\%    \\
                                         & ST & 9\%  & 35\%  & 70\% & 84\% \\ \hline
            \multirow{2}{*}{t$_{L}$}     & FD & 87\% & 78\%  & 66\% & 35\% \\
                                         & ST & 3\%  & 5\%   & 17\% & 38\% \\ \hline
            \multirow{2}{*}{t$_{M}$}     & FD & 11\% & 34\%  & 25\% & 13\%  \\
                                         & ST & 13\% & 35\%  & 54\% & 65\%  \\ \hline
            \multirow{2}{*}{Z$_{L}$}     & FD & 2\%  & 7\%   & 39\% & 38\% \\
                                         & ST & 2\%  & 12\%  & 12\% & 13\%  \\ \hline
            \multirow{2}{*}{Z$_{M}$}     & FD & 0\%  & 13\%  & 24\% & 31\% \\
                                         & ST & 10\% & 16\%  & 16\% & 18\% \\ \hline
            \multicolumn{2}{c|}{Total Number} & 87        & 171        & 187        & 55       
        \end{tabular}
        \tablefoot{We present the percentage of galaxies for which either FADO or STARLIGHT obtains an estimate which is significantly different relative to the estimate of the other code, considering an error uncertainty of 0.2 dex. We divide the analysis in four intervals of EW(H$\alpha$), in angstroms, and provide the number of objects in each interval.}
        \label{tab:FD_vs_ST}
    \end{table*}


\section{Discussion}
\label{discussion}

\begin{figure}[h!]
        \centering
        \includegraphics[scale=0.43]{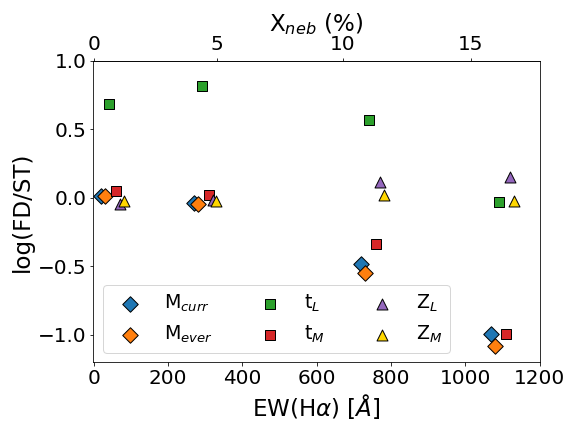}
        \caption{Logarithm of the median differences between FADO (FD) and STARLIGHT (ST) estimates of the stellar mass (diamonds), age (squares) and metallicity (triangles) for the previously considered EW(H$\alpha$) bins. Points in each bin slightly shifted for clearer view.}
        \label{fig:FD_vs_ST_vs_Xneb}
    \end{figure}

With the determination of the best tracers of the nebular contribution and the study of the impact of considering or not the nebular emission for different physical properties of galaxies, we can now evaluate the importance of taking into account the nebular emission depending on its contribution to the total emission. Based on these results, we discuss the implications for future analysis of both low- and high-redshift galaxies. In Fig. \ref{fig:FD_vs_ST_vs_Xneb}, the logarithm of the median differences between FADO and STARLIGHT estimates of the stellar mass, age and metallicity for the considered four EW(H$\alpha$) bins are shown as function of EW(H$\alpha$) and X$_{neb}$, thus summarising the results presented in Figs. \ref{fig:FD_vs_ST_M}, \ref{fig:FD_vs_ST_t}, and \ref{fig:FD_vs_ST_Z}.

\subsection{The importance of including nebular emission in spectral modelling}

From our comparison between FADO and STARLIGHT, we observe that for both currently available and total ever formed stellar mass, STARLIGHT tends to obtain larger estimates. This effect becomes increasingly more significant with the increase of the nebular contribution and is most noticeable when EW(H$\alpha$)$\geq$500 \AA, where for more than 70\% of our galaxies, considering the nebular emission leads to lower stellar mass estimates, with an average difference above 0.6 dex. As the EW(H$\alpha$) increases beyond this threshold, the differences can reach up to 2 dex. Thus, we can identify the value EW(H$\alpha$)=500 \AA\ as a clear threshold for which the neglect of nebular emission in spectral models leads to significant differences in the derived stellar masses.

For the light-weighted stellar age, we find that for EW(H$\alpha$)$<$1000 \AA, FADO tends to obtain larger estimates relative to STARLIGHT, while for EW(H$\alpha$)$\geq$1000 \AA\ the estimates are compatible. Looking to the mass-weighted stellar age, the results show that the differences between the estimates of the two codes become significant for EW(H$\alpha$)$\geq$1000 \AA, with STARLIGHT obtaining larger estimates. For lower values, there is also a slight tendency for STARLIGHT to obtain larger estimates.

In this way, we can establish EW(H$\alpha$)=1000 \AA\ as a threshold for clear differences in the estimation of the stellar age due to neglecting the nebular emission. However, since considering the nebular emission impacts stellar mass estimates for EW(H$\alpha$)$\geq$500 \AA, then the derived stellar ages could also be impacted at this level, due to a change in the fitted stellar populations. We will return to this point and further discuss it.

Relative to the light- and mass-weighted metallicity, our analysis suggests that there is no clear difference between the estimates of the two codes. Even if FADO has a tendency to estimate higher values for EW(H$\alpha$)$\geq$500 \AA, these differences never become so significant that a clear impact due to considering the nebular emission can be identified. Nonetheless, since for EW(H$\alpha$)$\geq$500 \AA\ the stellar mass estimates are impacted as consequence of accounting for the nebular emission, which in turn should impact the stellar age estimates, due to a change in the fitted stellar populations, as previously mentioned, then this change in stellar populations should also impact the estimated stellar metallicities. Moreover, the well-known degeneracy between stellar age and metallicity further suggests that above this limit, the stellar metallicity estimates must be affected as result of taking into consideration the nebular emission \citep[see, for example,][for a discussion of the influence of the age-metallicity degeneracy in spectral fitting]{con13}.

The evolution of the difference between FADO and STARLIGHT estimates with varying level of nebular contribution that we observe in this work is qualitatively similar to the results from \cite{FADO}, where FADO and STARLIGHT were applied to synthetic spectra. Our results can be understood as a consequence of the additional consideration of the nebular emission. The impact of including the nebular emission when modelling the optical spectrum of a galaxy has two main origins: 1) the relative contribution of the nebular to the total continuum and 2) the flat spectral shape of the nebular continuum (in the wavelengths between the Balmer and Paschen discontinuities). Both these points can be seen in Fig. \ref{fig:Xneb_vs_lambda}. Regarding the first point, by dividing the total continuum in the contribution from the nebular and stellar continuum, then the stellar continuum will be at a lower level relative to when the nebular continuum is neglected, since in the latter case the total continuum corresponds to the stellar continuum. In relation to the second point, since the nebular continuum is flat, a spectral fitting tool that does not take it into consideration will select an older stellar population in order to account for the flat continuum. Such nebular continuum modelling effects have been pointed out in several works \citep[e.g.][]{izo11,FADO,breda22,card22,mir23}.

Considering these effects, the inclusion of the nebular emission leads to a clear impact on the stellar mass estimates. As mentioned in the previous paragraph, considering both stellar and nebular emission leads to a lower stellar continuum and the selection of younger stellar populations, which have lower mass-to-light ratios. Thus, for galaxies where the nebular emission is significant, then the stellar mass estimates will be lower.

On the other hand, for the stellar age and metallicity the interpretation is more complex. The previously mentioned impacts on the stellar continuum also imply changes on the fitted stellar populations, which in turn will lead to different estimates of the stellar age and metallicity. As a matter of fact, this effect was studied by \cite{card22} where it was shown that considering the nebular emission leads to the estimation of higher light-weighted stellar ages, but compatible mass-weighted stellar ages. This was attributed to differences in the light and mass contributions of stellar populations younger than 1 Gyr as consequence of taking into account the nebular emission. For the light- and mass-weighted stellar metallicities, analogous results were observed. These results highlight that the extra inclusion of the nebular emission leads to different stellar populations being employed to fit the observed spectrum of a galaxy. Hence, it can be inferred that changes in the fitted stellar populations are one of the factors driving the differences in the stellar age and metallicity estimates as consequence of also considering the nebular emission.

In contrast with the results from \cite{card22}, we observe that considering the nebular emission can also significantly impact the mass-weighted stellar age estimates. This difference is related to the fact that in our work we observe these deviations for galaxies with EW(H$\alpha$)$\geq$1000 \AA, whereas in \cite{card22} the highest EW(H$\alpha$) interval is EW(H$\alpha$)$\geq$75 \AA, containing very few galaxies with EW(H$\alpha$)$\geq$150 \AA. In fact, if we consider our EW(H$\alpha$)$<$100 \AA\ interval, we see that our results are comparable with the ones from \cite{card22}. This difference in behaviour that arises for the galaxies with the highest nebular contributions means that the nebular contribution can reach such a level that the impact on the selected stellar populations is dramatic, thus influencing both light- and mass-weighted physical properties.

The discussion we undertake in this section is affected by two important factors that are worth discussing further: (1) the age-metallicity degeneracy and (2) the intrinsic differences between the two considered codes, FADO and STARLIGHT. The age-metallicity degeneracy relates to the fact that both old and metal rich stellar populations impact the spectra in similar ways, namely they tend to redden the spectra. With respect to the second point, it has been highlighted that the inner workings of different spectral synthesis tools might have a non-negligible effect on the derived physical properties \citep{breda22,card22,mir23}. Specifically, the minimisation algorithm used to find the best fitting model could play a major role in differences between spectral fitting codes.

It is foreseeable that our results are impacted by an interplay of these two factors, specifically the different codes can reach different best-fitting models due to their different fitting algorithms and this effect can be exacerbated by the age-metallicity degeneracy. For example, one of the codes could select a set of older stellar populations, thus leading to an older mean stellar age, whereas the other code could select a set of stellar population with higher metallicity, leading to a higher mean stellar metallicity. In the end, both fits can be equally good in terms of their $\chi^2$, but lead to different estimates of the physical properties. Despite these difficulties, by fixing the variables related to the spectral fitting and conducting a judicious analysis of the results, namely considering a conservative error interval for the physical parameters, it is still achievable to derive conclusions.

As a consistency check of our results, we made use of the capability of FADO to be used in pure stellar mode. By running FADO in pure-stellar mode (PS) and comparing the results to the ones previously obtained for the full-consistency mode (FC), we can remove from the analysis the uncertainties related to the intrinsic differences between FADO and STARLIGHT. Furthermore, with this comparison we can also establish if the inclusion of the nebular emission is in fact the main driver of the differences obtained in Sect. \ref{results}.

\begin{figure}[h!]
        \centering
        \includegraphics[scale=0.43]{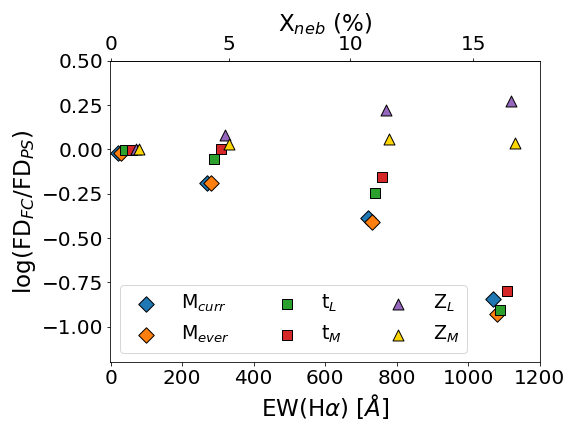}
        \caption{Logarithm of the median differences between FADO in full-consistency mode (FD$_{FC}$) and in pure-stellar mode (FD$_{PS}$) estimates of the stellar mass (diamonds), age (squares) and metallicity (triangles) for the previously considered EW(H$\alpha$) bins. Points in each bin slightly shifted for clearer view.}
        \label{fig:FD-FC_vs_ST_vs_Xneb}
    \end{figure}

To do this test, we applied FADO to our sample with the same parameters as the ones described in Sect. \ref{method}, but this time in PS mode. The results are summarised in Fig. \ref{fig:FD-FC_vs_ST_vs_Xneb}, where we compare the estimates of the considered physical properties of galaxies obtained by FADO in full-consistency mode (FD$_{FC}$) and in pure-stellar mode (FD$_{PS}$). The full distribution of the physical properties investigated in Fig. \ref{fig:FD-FC_vs_ST_vs_Xneb} are reported in the appendix (Figs. \ref{fig:FD_FC_vs_PS_M}, \ref{fig:FD_FC_vs_PS_t}, and \ref{fig:FD_FC_vs_PS_Z}).

There is a trend for the differences between the FC and PS modes to be mainly within the uncertainties when considering the lower EW(H$\alpha$) bins, but they increase and become significant when EW(H$\alpha$)$\geq$500 \AA. The stellar mass and age tend to be overestimated due to not considering the nebular emission, reaching $\sim$0.75 dex differences for EW(H$\alpha$)$\geq$1000 \AA. In contrast, the stellar metallicity tends to be underestimated when neglecting the nebular emission, although at a lower level when compared to the other physical properties.

Furthermore, by comparing Figs. \ref{fig:FD_vs_ST_vs_Xneb} and \ref{fig:FD-FC_vs_ST_vs_Xneb} we can have hints regarding the impact of the intrinsic differences between the two codes. We see that there is relatively good agreement between the two figures for all the properties except for the light-weighted stellar age. In this case, for the comparison between FADO and STARLIGHT, FADO retrieves higher estimates relatively to STARLIGHT and the difference decreases with increasing EW(H$\alpha$) and are negligible for EW(H$\alpha$)$\geq$1000 \AA. On the other hand, for the comparison between FADO in FC and PS mode, there is a negligible difference between the estimates for the lower EW(H$\alpha$) bins that becomes significant  with increasing EW(H$\alpha$), towards FADO in PS mode obtaining higher estimates. This points to some internal difference between FADO and STARLIGHT that particularly affects the light-weighted stellar age.

This analysis shows that, even if intrinsic differences between the two codes play a role, particularly in the light-weighted stellar age estimates, the main conclusions about the relevance of considering the nebular emission when fitting the spectra and deriving galaxy properties remain valid. Moreover, the previously identified EW(H$\alpha$)$\geq$500 \AA\ threshold also remains applicable.

Summarising, the most evident impact of considering the nebular emission is on the derived stellar masses, leading to lower estimates. Nevertheless, an impact on the derived stellar ages and metallicities can also be inferred, considering the implications on the fitted stellar populations of a stellar spectrum at different levels and with different shapes. Thus, we reach the conclusion that for galaxies with EW(H$\alpha$)$\geq$500 \AA\ the impact of considering the nebular emission is clear and significant. Considering Eq. \ref{eq:Xneb_Ha}, this threshold corresponds to X$_{neb}$$\sim$8\%, meaning that for galaxies with a median nebular contribution above this value it is fundamental to model both stellar and nebular emission. Using Eq. \ref{eq:Xneb_Hb} and \ref{eq:Xneb_sSFR}, we obtain the threshold in terms of EW(H$\beta$) and sSFR: EW(H$\beta$)$\simeq$110 \AA\ and $\log$(sSFR [yr$^{-1}$])$\simeq$-7.8, respectively. However, we note that it is possible that galaxies with median nebular contributions lower than 8\% are impacted by the inclusion of the nebular emission, but this is at a level which is more subtle and difficult to assess.

Considering the different method used by FADO to measure the EW (see Sect. \ref{method} for details), then we need to scale our threshold so that it is comparable to works that calculate the EW based on a pseudo-continuum estimate. Thus, the EW(H$\alpha$)=500 \AA\, threshold defined in this work corresponds to an EW(H$\alpha$)$\simeq$375 \AA\ threshold for works that use this different EW definition.

\subsection{Potential impact at low and high-redshifts}

Considering X$_{neb}$=8\% as the threshold above which there is an impact on the derived physical properties of galaxies, it is interesting to discuss its relevance for optical spectroscopic studies. Given the aforementioned evolution of the SF activity throughout the history of the Universe, considering the nebular contribution will have a different impact depending on the considered epoch.

At low redshifts, it is not expected that the nebular emission will have an impact for most galaxies, as explained in Sect. \ref{introduction}. Revisiting the results from \cite{mir23}, they observed no significant impact in the SFMS due to the extra inclusion of the nebular emission and this result was attributed to the fact that most galaxies in the sample did not host sufficiently high SF activity in order to have a significant nebular contribution. Their sample consists of $\sim$180 000 SF galaxies from SDSS-DR7, with a median EW(H$\alpha$)=34 \AA\ and 6\% of the sample has EW(H$\alpha$)$>$100 \AA, considerably below the EW(H$\alpha$)$\geq$500 \AA\ threshold defined in this work. In this way, we can confirm their hypothesis that for galaxies belonging to the SFMS at low redshift, the nebular contribution is not high enough so that considering it impacts significantly the derived physical properties of galaxies.

On the other hand, \cite{breda22} observed significant differences in the estimated physical properties for a sample of EELGs due to the considering the nebular emission. The EW(H$\alpha$) distribution of their sample has an average value of $\sim$400 \AA, with approximately half of the sample having EW(H$\alpha$)$\gtrsim$500 \AA. Therefore, at least 50\% of the galaxies in their sample have an EW(H$\alpha$) above the threshold defined in this work and the nebular emission should not be neglected.

Thus, our results are aligned with previous studies of the subject and highlight that while for most SF galaxies at low redshift the nebular emission is negligible, there is a subset for which it is fundamental to model the nebular emission to obtain accurate estimates of the physical properties. Amongst this subset of galaxies, it is worth highlighting the extremely metal-poor EELGs, often regarded as local analogues of high-redshift galaxies. Although rare at low redshifts, recent works have proposed that they are much more numerous at higher redshifts and likely to be common during the Epoch of Reionisation \citep{end23,boy24}. Considering the importance of these objects in understanding galaxy evolution in the young Universe and their role in the reionisation epoch, it is fundamental to accurately characterise their properties and to do that it is fundamental to carry out an adequate modelling of both stellar and nebular emission.

\begin{figure}[h!]
        \centering
        \includegraphics[scale=0.35]{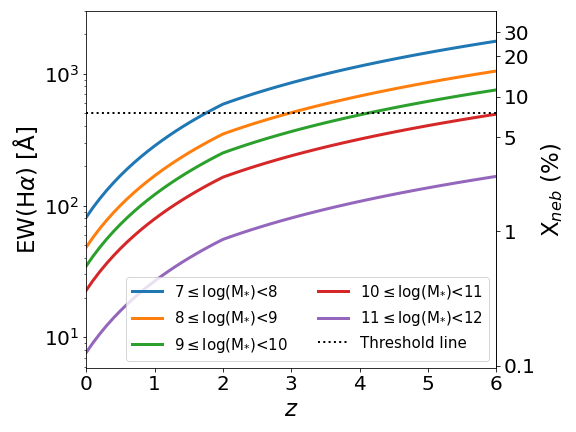}
        \caption{Evolution of the rest-frame EW(H$\alpha$) with redshift for different stellar mass intervals. We considered the EW(H$\alpha$) redshift evolution from \cite{fais16}. We also present, on the right-side axis, the median optical nebular contribution corresponding to the EW(H$\alpha$), calculated using the relation derived in this work. The black dotted line is the EW(H$\alpha$)=500 \AA\ threshold defined in this work.}
        \label{fig:EW_Halpha_vs_z}
    \end{figure}

As discussed in Sect. \ref{introduction}, the general level of SF activity of the Universe was higher in the past. Considering specifically the evolution of EW(H$\alpha$) with redshift, we can infer how the nebular contribution evolves with redshift, since we related the two quantities in Sect \ref{Neb_cont}. In \cite{fais16}, the evolution with redshift of the rest-frame EW(H$\alpha$) was studied for a sample with a median stellar mass of log(M$_{*}$/M$_{\odot}$)$\sim$9.8, obtaining the following empirical relation: EW(H$\alpha$) $\propto$ (1 + z)$^{1.8}$ at z$<$2 and $\propto$ (1 + z)$^{1.3}$ at 2$\leq$z$<$6. Figure \ref{fig:EW_Halpha_vs_z} shows the redshift evolution of the rest-frame EW(H$\alpha$) for different stellar mass intervals following the empirical relation derived in \cite{fais16}. We used the data from \cite{card22} to normalise the EW(H$\alpha$) evolution with redshift. We started by selecting only SF galaxies and then dividing them in bins of stellar mass. After that, for each stellar mass bin, we calculated the average EW(H$\alpha$) and redshift. These values are used to parameterise the redshift evolution of EW(H$\alpha$) for each stellar mass bin.

From Fig. \ref{fig:EW_Halpha_vs_z}, we can estimate the average EW(H$\alpha$) for a galaxy with a specific stellar mass as a function of redshfit. Hence, we can also calculate at what redshift a galaxy with a specific stellar mass will have, on average, an EW(H$\alpha$) above the defined threshold of 500 \AA. This will mean that the derived physical properties of that galaxy are significantly impacted due to considering both stellar and nebular emission.

Since we consider the same evolution of EW(H$\alpha$) with redshift regardless of the stellar mass, the redshift at which the threshold is crossed depends on the average EW(H$\alpha$) at z$\sim$0. Hence, the lower the stellar mass of the galaxy, the lower the redshift at which the threshold is crossed. Starting with the highest stellar mass interval, we observe that it does not reach the threshold in the considered redshift range. For the other stellar mass intervals, between log(M$_{*}$ [M$_{\odot}$])=7--11 in steps of 1 dex, the threshold is reached between $z$$\sim$2--6. This means that galaxies in the stellar mass range between log(M$_{*}$ [M$_{\odot}$])=7--11 will reach a point in the $z$$\sim$2--6 interval, where on average the nebular contribution in the optical is so significant that it cannot be neglected when carrying out spectral fitting. However, it is important to note that these results only hold on average and do not exclude the possibility that a galaxy of any mass experiencing an intense episode of SF activity surpasses the defined EW(H$\alpha$) threshold, thus leading to the necessity of accounting for both stellar and nebular emission.

Considering again the specific case of EELGs, works by \cite{boy24} and \cite{ller24} have conducted a census and characterisation of these objects at high-redshift ($z$$\sim$3--9) using data from JADES \citep{bunk20,riek20,eis23} and CEERS \citep{fink23,fink24}, respectively. In these works, the EW(H$\alpha$) of the EELGs ranges mainly between 300--1500 \AA, but goes up to $\sim$2000 \AA. Consequently, for almost all these galaxies, the optical nebular emission is so significant that it must be properly accounted for in order to obtain well determined physical properties.

With the extraordinary capabilities of JWST, an increasing number of objects are being found at these redshifts and even higher. In the near future these numbers will keep increasing with the onset of new high resolution spectroscopy instruments, namely MOONS. This means that for a progressively higher number of galaxies, considering both stellar and nebular emission in a self-consistent way is fundamental to accurately recover their physical properties and thus further develop our understanding of the process through which galaxies evolve in the young Universe.

\section{Summary and conclusions}
\label{conclusion}

The contribution of the nebular to the total optical emission has been shown to be highly relevant when modelling galaxies in periods of intense SF activity. At low redshifts, only a small fraction of galaxies host a sufficiently large SF activity to have a significant nebular contribution, the EELGs. In contrast, the increase in the general SF level of galaxies at higher redshifts hints that the nebular contribution should be relevant to take into consideration for a larger amount of galaxies. However, a threshold at which it becomes necessary to model the optical nebular contribution  has not been established. Our goal is to define this threshold and discuss the implications for both low- and high-redshift galaxies.

We selected a sample of 500 SF galaxies from SDSS-DR7, spanning a varying level of SF activity based on the EW(H$\alpha$) (see Fig. \ref{fig:Dist_Halpha}). Next, we applied to our sample FADO, a code that self-consistently fits the stellar and nebular continuum of the optical spectrum of a galaxy, and STARLIGHT, another population synthesis code, which lacks self-consistency prescriptions and was run with a purely stellar base. With the stellar and nebular continuum fitted by FADO we estimate the median optical nebular contribution, X$_{neb}$, for our sample (see Fig. \ref{fig:Xneb_dist}). With estimates of the physical properties of both codes, we study how they differ between each other depending on the median nebular contribution.

We started by evaluating different tracers of X$_{neb}$: the EW of H$\alpha$ and H$\beta$, the sum of the fluxes of the most prominent optical emission lines, SFR and sSFR (see Fig. \ref{fig:Xneb_vs_tracers_fit}). We find that the EW of H$\alpha$ and H$\beta$ are good tracers of X$_{neb}$, having a tight positive linear correlation. The sSFR could also be used as an approximate tracer of X$_{neb}$ as the relation is relatively tight; however, it is subject to caveats that need to be carefully taken into account. On the other hand, the sum of the fluxes of the most prominent optical emission lines and the SFR are not good tracers; despite showing a positive correlation with X$_{neb}$, the scatter of the relation is substantial. We also expanded this analysis to other emission lines, finding that the EW of [OIII]$\lambda$5007 and HeI$\lambda$5876 can also serve as possible tracers of X$_{neb}$ (see Fig. \ref{fig:Xneb_vs_tracers2_fit}).

Using the EW(H$\alpha$), we divided the sample in four intervals and studied how the differences between FADO and STARLIGHT (when estimating the physical properties of galaxies) are impacted with an increase in the value of EW(H$\alpha$) and, consequently, in the median nebular contribution (see Figs. \ref{fig:FD_vs_ST_M}, \ref{fig:FD_vs_ST_t}, and \ref{fig:FD_vs_ST_Z} and Table \ref{tab:FD_vs_ST}). Our results show that the self-consistent consideration of the nebular emission has a significant impact on the physical properties of galaxies when X$_{neb}$$\gtrsim$8\%, corresponding to EW(H$\alpha$)$\simeq$500 \AA, EW(H$\beta$)$\simeq$110 \AA\ and sSFR$\simeq$10$^{-7.8}$ yr$^{-1}$. For the EW of [OIII]$\lambda$5007 and HeI$\lambda$5876, this corresponds to $\sim$390 \AA\, and $\sim$14 \AA, respectively. Given the different definition of FADO for the continuum, for works considering a pseudo-continuum, these EW values actually correspond to: EW(H$\alpha$)$\simeq$375 \AA, EW(H$\beta$)$\simeq$80 \AA, EW([OIII]$\lambda$5007)$\simeq$290 \AA\ and EW(HeI$\lambda$5876)$\simeq$10 \AA.

At the defined threshold, taking into account both stellar and nebular emission has a clear impact on the stellar mass, leading to lower estimates for more than 70\% of the galaxies with an average 0.6 dex difference, rising up to 2 dex for the galaxies with the highest nebular contribution. For the stellar age and metallicity, the differences are not as straightforward as for the stellar mass. For the light-weighted stellar age, for X$_{neb}$$\lesssim$15\% (EW(H$\alpha$)$\simeq$1000 \AA) there is a trend for obtaining older ages when considering the nebular emission, whereas for the mass-weighted  stellar ages only for galaxies with X$_{neb}$$\gtrsim$15\% is there a significant difference when including the nebular emission, leading to younger ages. In the case of light- and mass-weighted stellar metallicities there is no clear difference on the estimates due to accounting for both stellar and nebular emission. However, the inclusion of the nebular emission and its subsequent impact on the stellar mass estimate, also affects the estimated stellar populations that constitute the galaxy. Thus, even though the consequence of taking into account the nebular emission on the stellar age and metallicity is not easily observable, it can be inferred from our results due to a change in the underlying fitted stellar populations of the galaxy.

With the aim of testing if the intrinsic difference between FADO and STARLIGHT has an impact in these results, we applied FADO in pure-stellar mode to our sample and compared to the estimates previously obtained (see Fig. \ref{fig:FD-FC_vs_ST_vs_Xneb}). This analysis has demonstrated that the intrinsic differences between the two codes play a role, but the derived results regarding the importance of considering the nebular emission when fitting the spectra and deriving galaxy properties (including the thresholds mentioned above) remain valid.

Our results imply that at low redshifts, where the nebular contribution is low, most galaxies are not impacted by also considering the nebular emission. However, there is a subset of galaxies in a phase of very strong SF activity, the EELGs, for which the contribution of the nebular emission is non-negligible and impacts the inferred physical properties and stellar populations that compose the galaxy. At higher redshifts, there are several evidences for an increase of the general SF activity level of galaxies. Considering the EW(H$\alpha$) evolution with redshift, we find that galaxies with M$_{*}$=10$^{7}$--10$^{11}$ M$_{\odot}$ will, on average, reach and surpass the defined threshold at $z$$\sim$2--6 (see Fig. \ref{fig:EW_Halpha_vs_z}). This implies that considering the nebular emission is a vital aspect for most galaxies at these redshifts. 

This work shows that accounting for both the optical stellar and nebular continuum is fundamental for EELGs at low redshifts and more generally for SF galaxies at high-redshifts, specifically at z$>$2. In light of the recent JWST observations of high-redshift galaxies and future instruments aimed at exploring the young Universe (e.g. MOONS) this calls attention to a crucial aspect to take into consideration when analysing the physical and evolutionary properties of galaxies. In the future, this work should be expanded towards high-redshift galaxies to directly estimate their nebular contribution and its relevance in spectral fitting. Additionally, other commonly used spectral fitting tools that include nebular emission in their models, such as Prospector \citep{leja17,john21} and Bagpipes \citep{carn18}, should be included in upcoming studies of the nebular contribution of high-redshift galaxies.

\begin{acknowledgements}

This work was supported by Fundação para a Ciência e a Tecnologia (FCT) through the research grants PTDC/FIS-AST/29245/2017, UID/FIS/04434/2019, UIDB/04434/2020 and UIDP/04434/2020.  H.M. acknowledges support from the Fundação para a Ciência e a Tecnologia (FCT) through the PhD Fellowship 2022.12891.BD. C. P. acknowledges support from DL 57/2016 (P2460) from the ‘Departamento de Física, Faculdade de Ciências da Universidade de Lisboa’. P.P. acknowledges support by FCT through  Principal Investigator contract CIAAUP-092023-CTTI. R.C. acknowledges support from FCT through the Fellowship PD/BD/150455/2019 (PhD:SPACE Doctoral Network PD/00040/2012) and POCH/FSE (EC). P.L. (DOI 10.54499/dl57/2016/CP1364/CT0010) is supported by national funds through Fundação para a Ciência e a Tecnologia (FCT) and the Centro de Astrofísica da Universidade do Porto (CAUP). D.B. acknowledges support from FCT through the Fellowship UI/BD/152315/2021. Funding for the SDSS and SDSS-II has been provided by the Alfred P. Sloan Foundation, the Participating Institutions, the National Science Foundation, the U.S. Department of Energy, the National Aeronautics and Space Administration, the Japanese Monbukagakusho, the Max Planck Society, and the Higher Education Funding Council for England. The SDSS Web Site is \url{http://www.sdss.org/}. The SDSS is managed by the Astrophysical Research Consortium for the Participating Institutions. The Participating Institutions are the American Museum of Natural History, Astrophysical Institute Potsdam, University of Basel, University of Cambridge, Case Western Reserve University, University of Chicago, Drexel University, Fermilab, the Institute for Advanced Study, the Japan Participation Group, Johns Hopkins University, the Joint Institute for Nuclear Astrophysics, the Kavli Institute for Particle Astrophysics and Cosmology, the Korean Scientist Group, the Chinese Academy of Sciences (LAMOST), Los Alamos National Laboratory, the Max-Planck-Institute for Astronomy (MPIA), the Max-Planck-Institute for Astrophysics (MPA), New Mexico State University, Ohio State University, University of Pittsburgh, University of Portsmouth, Princeton University, the United States Naval Observatory, and the University of Washington.

\end{acknowledgements}

\begin{appendix}
\label{appendix}
\section{Comparison of FADO estimates in pure-stellar and full-consistency mode}

\begin{figure}[h!]
        \centering
        \includegraphics[scale=0.3]{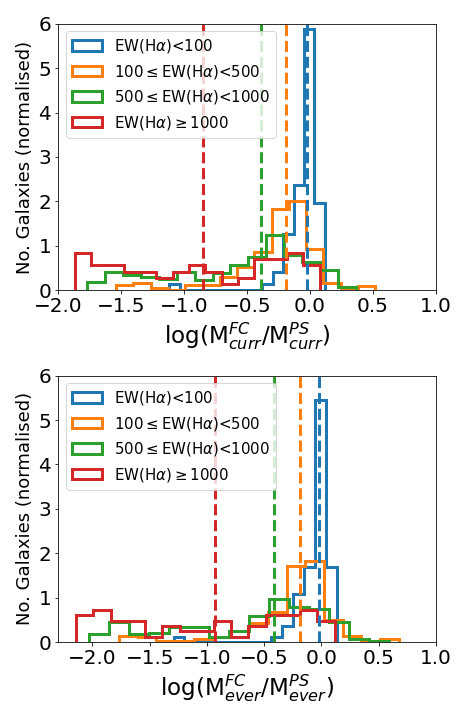}
        \caption{Distribution of the logarithmic difference between the currently available stellar mass (upper panel) and total ever formed stellar mass (bottom panel) estimated by FADO in full-consistency (FC) and pure-stellar (PS) mode, for four EW(H$\alpha$) bins: EW(H$\alpha$)$<$100 \AA\ (blue), 100$\leq$EW(H$\alpha$)$<$500 \AA\ (orange), 500$\leq$EW(H$\alpha$)$<$1000 \AA\ (green) and EW(H$\alpha$)$\geq$1000 \AA\ (red). The
vertical dashed lines represent the median value of each distribution.}
        \label{fig:FD_FC_vs_PS_M}
    \end{figure}

\begin{figure}[h!]
        \centering
        \includegraphics[scale=0.3]{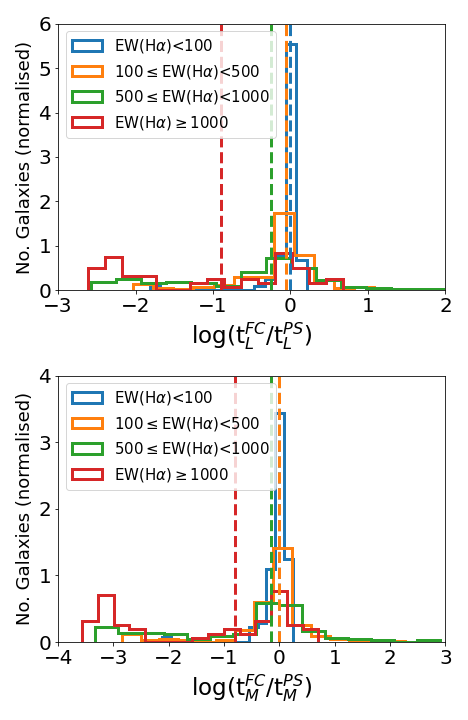}
        \caption{Same as Fig. \ref{fig:FD_FC_vs_PS_M}, but for the light- and mass-weighted stellar age (upper and lower panels, respectively).}
        \label{fig:FD_FC_vs_PS_t}
    \end{figure}

\begin{figure}[h!]
        \centering
        \includegraphics[scale=0.3]{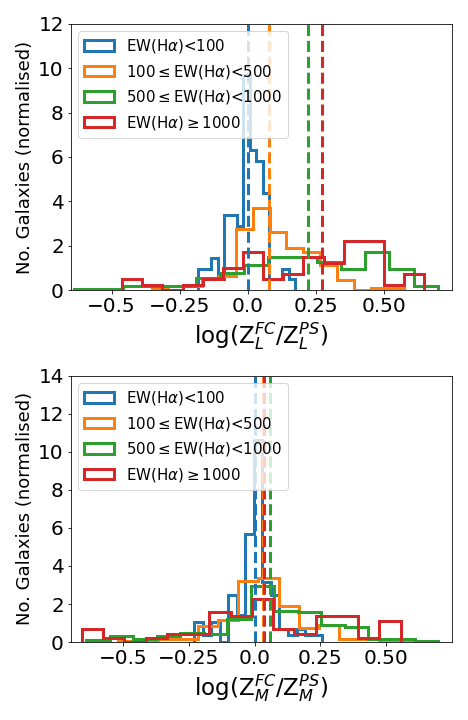}
        \caption{Same as Fig. \ref{fig:FD_FC_vs_PS_M}, but for the light- and mass-weighted stellar metallicity (upper and lower panels, respectively).}
        \label{fig:FD_FC_vs_PS_Z}
    \end{figure}

Figures \ref{fig:FD_FC_vs_PS_M}, \ref{fig:FD_FC_vs_PS_t}, and \ref{fig:FD_FC_vs_PS_Z} show the distribution of the logarithm of the ratio between the estimates obtained by FADO in full-consistency (FC) and pure-stellar (PS) mode of different physical properties of galaxies: stellar mass, age and metallicity, respectively. The distributions are normalised such that the area under the curve is equal to one. Additionally, in the plots, the positive values correspond to FC mode obtaining larger estimates relative to the PS mode and negative values correspond to the opposite case.

\end{appendix}

\end{document}